\documentclass{SciPost}

\usepackage{graphicx, siunitx, amssymb,amsmath,}

\begin{document}

\begin{center}{\Large \textbf{
Tunable dispersion of the edge states \\
in the integer quantum Hall effect
}}\end{center}

\begin{center}
M. Malki\textsuperscript{1*}, 
G.S. Uhrig\textsuperscript{1$^\diamond$}
\end{center}

\begin{center}
{\bf 1} Lehrstuhl f\"ur Theoretische Physik 1, TU Dortmund University, Germany
\\
* \href{mailto:maik.malki@tu-dortmund.de}{maik.malki@tu-dortmund.de} \\
$^\diamond$ \href{mailto:goetz.uhrig@tu-dortmund.de}{goetz.uhrig@tu-dortmund.de}
\end{center} 	

\begin{center}
\today
\end{center}


\section*{Abstract}
{\bf 
Topological aspects represent currently a boosting area in condensed matter physics.
Yet there are very few suggestions for technical applications of topological phenomena.
Still, the most important is the calibration of resistance standards by means of the
integer quantum Hall effect. We propose modifications of samples displaying 
the integer quantum Hall effect which render the tunability of the Fermi velocity 
possible by external control parameters such as gate voltages.
In this way, so far unexplored possibilities arise to realize devices such as 
tunable delay lines  and interferometers.
}

\vspace{10pt}
\noindent\rule{\textwidth}{1pt}
\tableofcontents\thispagestyle{fancy}
\noindent\rule{\textwidth}{1pt}
\vspace{10pt}

\section{Introduction}
\label{sec:intro}

\subsection{General context}

Subjecting a two-dimensional electron gas at low temperature to a strong perpendicular magnetic field results in the well-known quantization of the transverse conductivity 
$\sigma_{xy} = \nu e^2/ h$ with $\nu \in \mathbb{N}$ which is called integer quantum Hall effect  \cite{klitz80} (IQHE). The remarkable high-precision with which 
 the integer quantum Hall conductivity can be measured is attributed to its 
relation to topological invariants \cite{thoul82,avron83,niu85,kohmo85,hatsu93,altla10}. 
Shortly after the discovery of the IQHE another topological effect was measured and baptized 
the fractional quantum Hall effect  \cite{tsui82,laugh83} since Hall plateaus appear at fraction filling factors $\nu$. The discovery of the integer and fractional quantum Hall effect triggered
a steadily growing interest in topological phenomena in condensed matter phenomena.

The IQHE is a single-particle phenomenon \cite{laugh81,halpe82}; no interaction between the
electrons needs to be taken into account which facilitates its understanding greatly.
In the bulk, the interpretation of the IQHE is that
 the filling factor $\nu$ equals the total Chern number of the filled Landau bands.
This Chern number is a topological invariant \cite{avron83,niu85,kohmo85} related to 
the fundamental Berry phase \cite{berry84}. This warrants the high precision of resistance 
measurements fulfilling Ohm's law without any non-linear corrections \cite{uhrig91}.

A closer understanding is gained if one realizes that the 
actual charge currents are carried by gapless edge states \cite{hatsu93} which cross
the Fermi level. They have to exist 
at the boundaries because the Chern number jumps across them \cite{berne13}.
The number of gapless edge states \cite{thoul82} corresponds to the Chern number $\nu$. Each of 
these edge states can be seen as single-channel conductor \cite{butti85} propagating only in one
direction along the edge which therefore are called chiral edge states. They allow for adiabatic
transport \cite{beena91} because backscattering is forbidden which makes such transport
particularly interesting for applications. 

It is fascinating that the IQHE can be put into a larger context of Chern insulators \cite{berne13}
which need not be induced by external magnetic fields. Complex kinetic Hamilton
operators on lattices, i.e., complex hoppings, can imply non-trivial Chern numbers
and concomitant edge modes as they appear in the IQHE. The seminal example is the 
 Haldane model \cite{halda88b}. Its quantum Hall effect is called anomalous because
no magnetic field is required.
The inclusion of the spin degree of freedom \cite{kane05a,kane05b} opens the possibility of the quantum spin Hall effect \cite{weng15,liu16,ren16}. The quantum anomalous Hall effect 
\cite{chang13,kou14,chang15} as well as the quantum spin Hall effect \cite{ando13} have been realized experimentally.

\subsection{Present objective}

For clarity, we focus here on the IQHE and do not take the spin into account
which is left to future research. The topological
protection of the chiral edge states and the complete suppression of backscattering in these edge
states suggests that the chiral edge states enable robust applications. Calibrating
resistance standards to extremely high precision is certainly a wonderful example.
Yet, in the present study we want to trigger research on \emph{further} applications.

We will investigate the Fermi velocity $v_\mathrm{F}$ occurring in the chiral edge states.
It represents the group velocity of electrical signal transmitted through the system.
Hence, it determines the speed of signal transmission. If it can be tuned it can
be used to influence the time signals need to cross the sample. In this way, certain delays
can be imposed and used for signal processing, for instance for interference measurements. 
We emphasize that the Fermi velocity does not influence the widely studied 
DC conductivity which is not the quantity of interest in our study, in contrast to 
the majority of theoretical studies in the literature. 

Triggered by the observation that the Fermi velocity of edge states in Chern insulators
on lattices differs depending on the details of the edges \cite{redde16} 
a systematic study of modifications of the edges of the generic Chern insulator
in the Haldane model revealed that the Fermi velocity can indeed be tuned over
orders of magnitude by changing external parameters such as gate voltages \cite{uhrig16}.
The key idea is to modify the edges by decorations such that local levels are created which are
brought in weak contact with the dispersive edge modes. The ensuing hybridization
leads to a weakly dispersing mode of which the Fermi velocity can be tuned by
changing the energy of the local modes. If the local levels are in resonance with the edge modes
the sketched mechanism is at work and a low Fermi velocity appears.
If they are out-of-resonance the hybridization is ineffective and the 
edge states remain strongly dispersive. The tuning of the local decorated edge
modes can be achieved by gate voltages.

This fundamental idea has been carried over from the spinless Haldane model 
 to the spinful Kane-Mele model \cite{malki17b}. In this study, the effect of
disorder in the decorated Haldane model has been addressed as well.
It was shown that the Fermi velocity is robust against weak disorder
if the dispersion is not too flat, i.e., if the Fermi velocity is not
too low. Hence, in contrast to the naive expectation of complete robustness
due to the topological origin of the edge states disorder changes the dispersion
of the modes and can deteriorate signal transmission \emph{beyond}
the DC conductivity.

As pointed out in the general context, tunable Fermi velocities open the
possibility of interesting applications such as delay lines or interference
devices. Unfortunately, the lattice systems known so far cannot yet be tailored
on the nanoscale to render the experimental verification of the theoretical proposal
possible. So far, solid state systems postulated by density-functional theory
can be envisaged to yield realizations in the future \cite{liu11,wu14,han15}.
Alternatively, intricate optical lattice may make proof-of-principle
realizations of tunable Fermi velocities possible \cite{jotzu14,aidel15}.
Yet, the search for different realizations is called for.
In particular, the high standard of designing nanostructures in semiconductor
systems suggests to look for such systems for the realization of tunable
dispersions of edge states.

\begin{figure}[htb]
	\centering
		\includegraphics[width=\columnwidth]{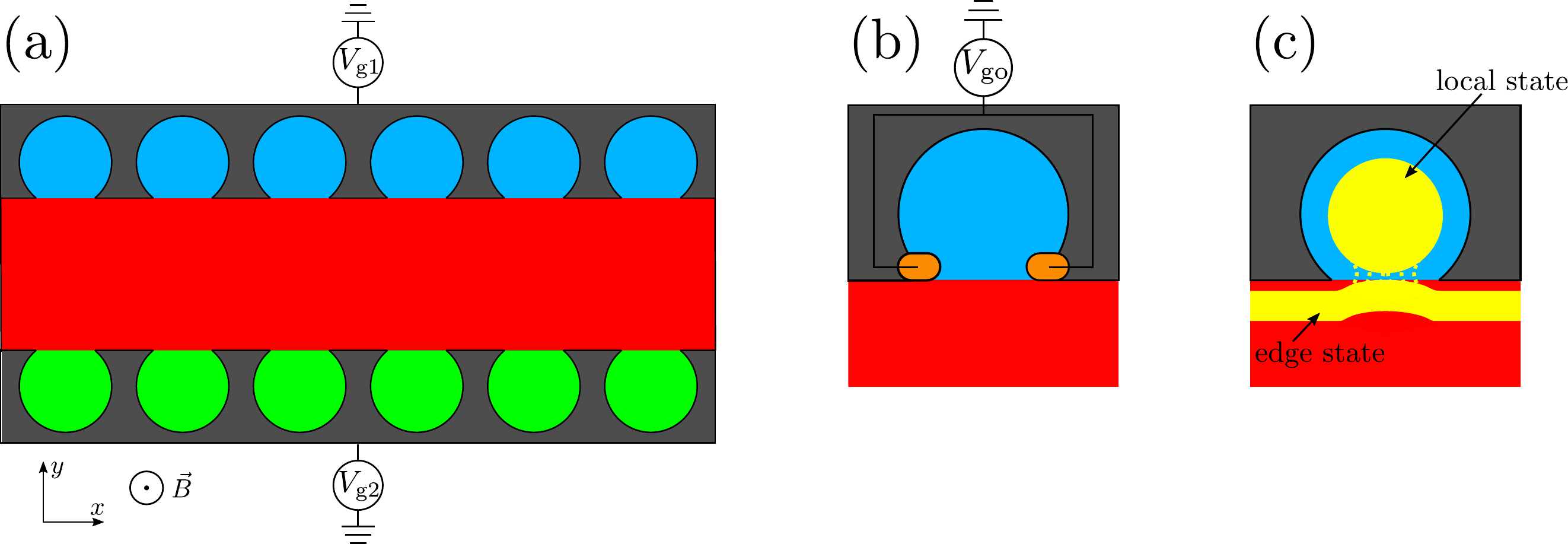} 
	\caption{Panel (a): proposal of a decorated quantum Hall sample with tunable Fermi velocity. A 
	perpendicular magnetic field puts the two-dimensional electron gas in the quantum 
	Hall phase. Two independent gate voltages $V_\text{g1}$ and $V_\text{g2}$
	change the potential of the blue bays at the upper boundary and of the green bays at the lower
	boundary, respectively. The grey area is inaccessible to the electrons.
		The size of the opening of the bays to the bulk 2DEG can be controlled
	by a gate voltage $V_\text{go}$ as depicted in panel (b). The size of the opening
	controls the degree of hybridization of the local mode within the bays and the
	edge mode in the 2D bulk, see panel (c).}
	\label{fig:sample}
\end{figure}

This brings us back to the IQHE which is based on a semiconducting
interface generating a two-dimensional (2D) electron gas and a perpendicular
magnetic field. If one is able to tailor the boundaries of the 2D electron gas
in a way that mimics the decoration of 2D lattice models tunable
Fermi velocities become possible. Indeed, it has been proposed by one
of us that attaching bays to the boundaries of a Hall sample allows
us to generate local modes in the bay \cite{uhrig16}. If they are slightly opened 
to the 2D bulk a weak hybridization is realized and the physics 
established so far for lattice systems should carry over to the
IQHE. The basic geometry is sketched in Fig.\ \ref{fig:sample}.

Currently, it is possible to implement bays in the submicrometer range in IQHE samples.
For instance, a single-electron source has been realized by coupling a quantum dot
to a 2DEG via quantum point contacts and a gate voltage setting the dot potential 
\cite{feve07}. An additional gate voltage at the quantum point contacts is used to control the transmission, see Fig.\ \ref{fig:sample}(b)
 so that the hybridization can be tuned as indicated in Fig.\ \ref{fig:sample}(c). 
If such a coupled quantum dot is repeated  periodically the geometry in Fig.\ \ref{fig:sample}(a)
is obtained. This proposed setup will be studied  in the sequel as 
an exemplary model for the realization of tunable Fermi velocities in the IQHE.

Below, we present calculations showing that the Fermi velocity $v_\mathrm{F}$ 
can be tuned by adding periodically arranged bays to an integer quantum Hall sample. 
The paper is organized as follows. In Sect.\ \ref{sec:model} we specify the model
Hamiltonian describing the IQHE and the numerical approach to 
compute the edge states and their dispersion.
Sect.\ \ref{sec:dispersion} illustrates step by step how the spectrum of the 
decorated IQHE is structured. In particular, we focus on the effects of
the hybridization between the modes in the bays and the edge modes because
this is the mechanism altering the  Fermi velocities.
The results for tuned Fermi velocities are presented in Sect.\ \ref{sec:tune}. 
Finally, Sect.\ \ref{sec:conclusion} collects our findings and provides an outlook.

\section{Model and technical aspects}
\label{sec:model}

\begin{figure}[htb]
	\centering
		\includegraphics[width=\columnwidth]{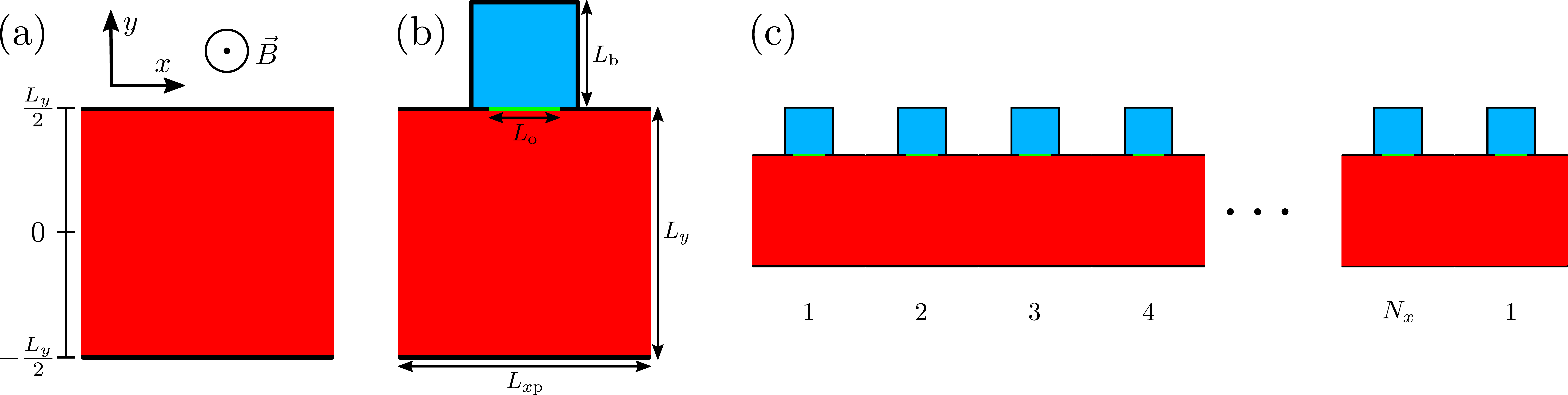} 
	\caption{Sketch of the considered geometries of increasing complexity. Panel (a) 
	displays the standard IQHE sample without any decoration of the boundaries; its width
	is denoted by $L_y$ and its total length by $L_x$. Periodic boundary conditions
	in $x$-direction are assumed. \smash{Panel (b)} shows the single unit considered where
	the dimensions of the bay and the coupled bulk are given. Note the opening of the
	bay shown in green; its width is denoted by $L_o$. The total sample consists
	of $N_x$ such units as shown in panel (c) so that $L_x=N_x L_{x\text{p}}$.}
	\label{fig:ez}
\end{figure}

The present work is designated to illustrate the tunability of the Fermi velocity
on a proof-of-principle level. For the sake of clarity, we assume that the upper and the
lower boundary are sufficiently far away from each other so that the edge states
localized at the upper and at the lower boundary do not influence each other.
Practically, this means that the magnetic length $\ell_B=\sqrt{\hbar/(|e B|)}$ 
is significantly smaller than
the width $L_y$ of the quantum Hall sample, i.e., the external magnetic field must
be large enough. Then, it is not necessary to study a system of which
both boundaries are decorated. Hence, we focus here on a sample with quadratic bays
at the upper boundaries, but no decoration at the lower boundary which 
is kept smooth. The precise shape of the bays does not matter for
our proof-of-principle calculations.
Within the colored area shown in the panels of Fig.\ \ref{fig:ez} the electrons
can move freely. Their dynamics is only governed by their kinetic energy. The boundaries
are supposed to be infinitely hard walls as indicated by thick black lines.

Applying a perpendicular magnetic field in $z$-direction, see panel (a) in Fig.\ \ref{fig:ez}, is
incorporated in the usual way by minimal coupling 
\begin{equation}
	\vec{p} \rightarrow \vec{p} - q \vec{A} 
\end{equation}
where the charge reads $q = - |e|$ and $\vec{A}$ is the magnetic vector potential.
No electron-electron interactions are considered so that the full Hamilton
operator reads
\begin{equation}
	H = \frac{1}{2m} \left( \vec{p} - q \vec{A} \right)^2
\end{equation}
where $m$ is the (effective) mass of the electrons. 
The electrons are confined to the  $xy$-plane; we do not consider their spin
degree of freedom. This can be justified because the two spin species $\uparrow$ and
$\downarrow$ are decoupled in the perpendicular magnetic field \cite{hwang93,parad10}. 

Due to the translational invariance in the $x$-direction a Landau gauge is particularly 
appropriate. We choose the Landau gauge in $x$-direction $\vec{A} = B (-y, 0, 0)$ 
so that the \smash{momentum  $k_x$} remains manifestly conserved. 
This leads to the continuum Hamilton operator
\begin{subequations}
\begin{align}
H_{\mathrm{bulk}} =& \frac{\hbar^2}{2m} \left[ \left( - \mathrm{i} \frac{\partial}{\partial x} + \frac{q B}{\hbar} y \right)^2 - \frac{\partial^2}{\partial y^2} \right]
\\
	=& \frac{m \omega_{\mathrm{c}}^2}{2} \left( y + \mathrm{i} \ell_B^2 \frac{\partial}{\partial x} \right)^2 - \frac{\hbar^2}{2 m} \frac{\partial^2}{\partial y^2} 
	\label{eq:landau}
\end{align}
\end{subequations}
 in real space where we use the definition of the cyclotron frequency 
$\omega_\mathrm{c} = |e|B/m$ and the magnetic length $\ell_B = \sqrt{\hbar/|eB|}$. 
It is implied that $x$ and $y$ take only values in the colored regions of the panels
in Fig.\ \ref{fig:ez} unless stated otherwise.

\subsection{Bulk system}
\label{sec:bulk}

Solving the Hamiltonian \eqref{eq:landau} in case of a bulk system 
without any boundaries leads to the famous Landau levels with quantized energy values 
\cite{landa30}
\begin{equation}
	E_n = \hbar \omega_\mathrm{c} \left(n + 1/2 \right), \ n \in \mathbb{N} \ .
	\label{eq:landau_energy}
\end{equation}
The corresponding wave functions are plane waves in
$x$-direction and Gaussians multiplied with Hermite polynomials in $y$-direction
\begin{equation}
	\psi(n, k_x, y) = 
	N \mathrm{e}^{-(y-y_0)^2/2 \ell_B^2} H_n((y-y_0)/\ell_B) \mathrm{e}^{\mathrm{i} k_x x}
	\label{eq:landau_wave}
\end{equation}
because the Hamiltonian corresponds to shifted harmonic oscillators in $y$-direction.
The wave functions are normalized by $N$, $H_n$ is the $n$th Hermite polynomial, and 
$y_0 = k_x \ell_B^2$ determines the center of the wave function $\psi(n, k_x, y)$ 
in $y$-direction. These facts about the bulk Landau level will be helpful for the 
 understanding of the more complicated situations and serve as reference.
Below, we consider more and more details of the actual model depicted in 
Fig.\ \ref{fig:ez}(c).

\subsection{Sample of finite width $L_y$}
\label{sec:hardwall}

Next, we consider a sample as shown in Fig.\ \ref{fig:ez}(a), i.e., of finite width
in $y$-direction, but with translational invariance along $x$ due to periodic boundary
conditions. A numerical treatment is required which we introduce here. It is chosen
flexible enough to be extended subsequently to the decorated sample including
the bays.

For simplicity we set the effective electron mass 
$m = 1$, Planck's constant $\hbar = 1$, and use $B$ henceforth for  $|e| B $. 
This amounts to setting $\omega_\text{c}=1$, i.e., to using $\hbar\omega_\text{c}$ as energy unit.
The resulting bulk Hamiltonian reads
\begin{equation}
	H_{\mathrm{bulk}} = \frac{1}{2} \left[ \left( \frac{y}{\ell_B^2}  + 
	\mathrm{i} \frac{\partial}{\partial x} \right)^2 - \frac{\partial^2}{\partial y^2} \right] .
\end{equation}
As displayed in Fig.\ \ref{fig:ez}(a) the boundary conditions in the $y$-direction 
imply $V(y) = \infty$ for $|y| \ge L_y/2$. We use the same Landau gauge as before in the bulk system. In $x$-direction, we exploit the translational invariance using the plane wave ansatz
\begin{equation}
\label{eq:plane}
\psi(x,y) = \exp(ik_x x) \psi(y).
\end{equation}
This leads to the Hamilton operator which acts on $\psi(y)$ 
\begin{equation}
	H_\mathrm{undec.~con.} = 
	\frac{1}{2} \left[ \left( \frac{y}{\ell_B^2}  - k_x \right)^2 - \frac{\partial^2}{\partial y^2} \right]
	\label{eq:H_cont_y}
\end{equation}
with $|y| \leq \frac{L_y}{2}$. We tackle this problem by discretizing the $y$ coordinate
by a mesh with \smash{distance $a$} between the points. It is understood that $a$ is much smaller
than any other physical length scale in the system, i.e., $\ell_B$ and $L_y$. 
The resulting model resembles a tight-binding model, but we emphasize that its discrete
character is just due to the approximate treatment of the continuum. We make sure
that the discretization mesh is always fine enough so that the results are close
to the continuum values, see below.

\begin{figure}[htb]
	\centering
		\includegraphics[width=0.5\columnwidth]{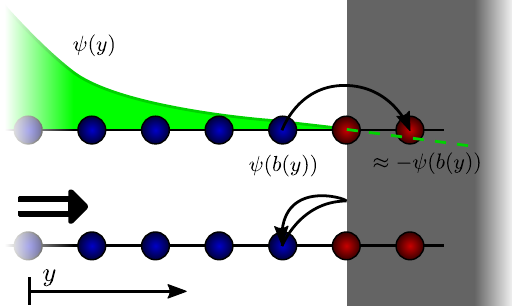} 
	\caption{Illustration of the approximation used in immediate vicinity of
	a boundary in order to improve the approximation of the continuous system
	by a discretized one, see main text.}
	\label{fig:spiegel}
\end{figure}

So the discretized Hamiltonian, expressed in second quantization, which approximates the continuum operator \eqref{eq:H_cont_y} reads
\begin{align}
 H_\mathrm{undec.~dis.} = \sum_y &\left[ \frac{1}{2} \left( \left( \frac{y}{\ell_B^2}  - k_x \right)^2 + \frac{5}{2 a^2} \right) c_{y, k_x}^{\dagger} c_{y, k_x} - \frac{2}{3 a^2} c_{y+a, k_x}^{\dagger} c_{y, k_x}
\right. \nonumber \\
 &  \left. + \frac{1}{24 a^2} c_{y+2a, k_x}^{\dagger} c_{y, k_x} + \mathrm{h.c.} \right] - \frac{1}{24 a^2} c_{b(y), k_x}^{\dagger} c_{b(y), k_x} 
\label{eq:boundaryterm}
\end{align}
where $c_{y, k_x}$ ($c_{y, k_x}^\dagger$) annihilates (creates) an electron with wave vector $k_x$
in $x$-direction at coordinate $y$. To this end, the second derivative is approximated
by the difference quotient
\begin{align}
  \frac{\partial^2 \psi(y)}{\partial y^2} &\approx 
	\frac{1}{a^2} \left[ -\frac{1}{12} \psi(y - 2a) + \frac{4}{3} \psi(y - a) 
	 - \frac{5}{2} \psi(y)  + \frac{4}{3} \psi(y + a) - \frac{1}{12}\psi(y + 2 a) \right] .
  \label{eq:differenz}
\end{align}
This formula cannot be applied to values of $y$ which are close to a boundary
because the values $\psi(y + a)$ and $\psi(y + 2a)$ may not exist, see Fig.\ \ref{fig:spiegel}.
In fact, if $y_\text{bdry}$ is the value right at the boundary 
(red site partly in the shaded area in 
Fig.\ \ref{fig:spiegel}), then $\psi(y_\text{bdry})=0$
holds due to the hard-wall boundary condition
and one does not need a terms at $y_\text{bdry}$. What is needed is an approximation of 
the second derivative at \smash{$y_\text{bdry}-a$} for which \smash{$\psi(y_\text{bdry}+a)$} 
is required. One could simply omit this term, but this omission would introduce 
an error of the order of $a$
with respect to the continuum situation which we intend to approximate. Hence, we exploit
that $\psi(y_\text{bdry})=0$ and that a continuous function can be approximated by its
Taylor expansion around $y_\text{bdry}$. In linear order this implies 
\smash{$\psi(y_\text{bdry}+a)\approx -\psi(y_\text{bdry}-a)$} which leads the last term in 
\eqref{eq:boundaryterm} where we used the symbol \smash{$b(y)=y_\text{bdry}-a$} 
for the value of $y$ adjacent to the boundary. This improves the results roughly by
one order in $a$, especially at the important edges of the sample.

\begin{figure}[htb]
	\centering
		\includegraphics[width=0.8\columnwidth]{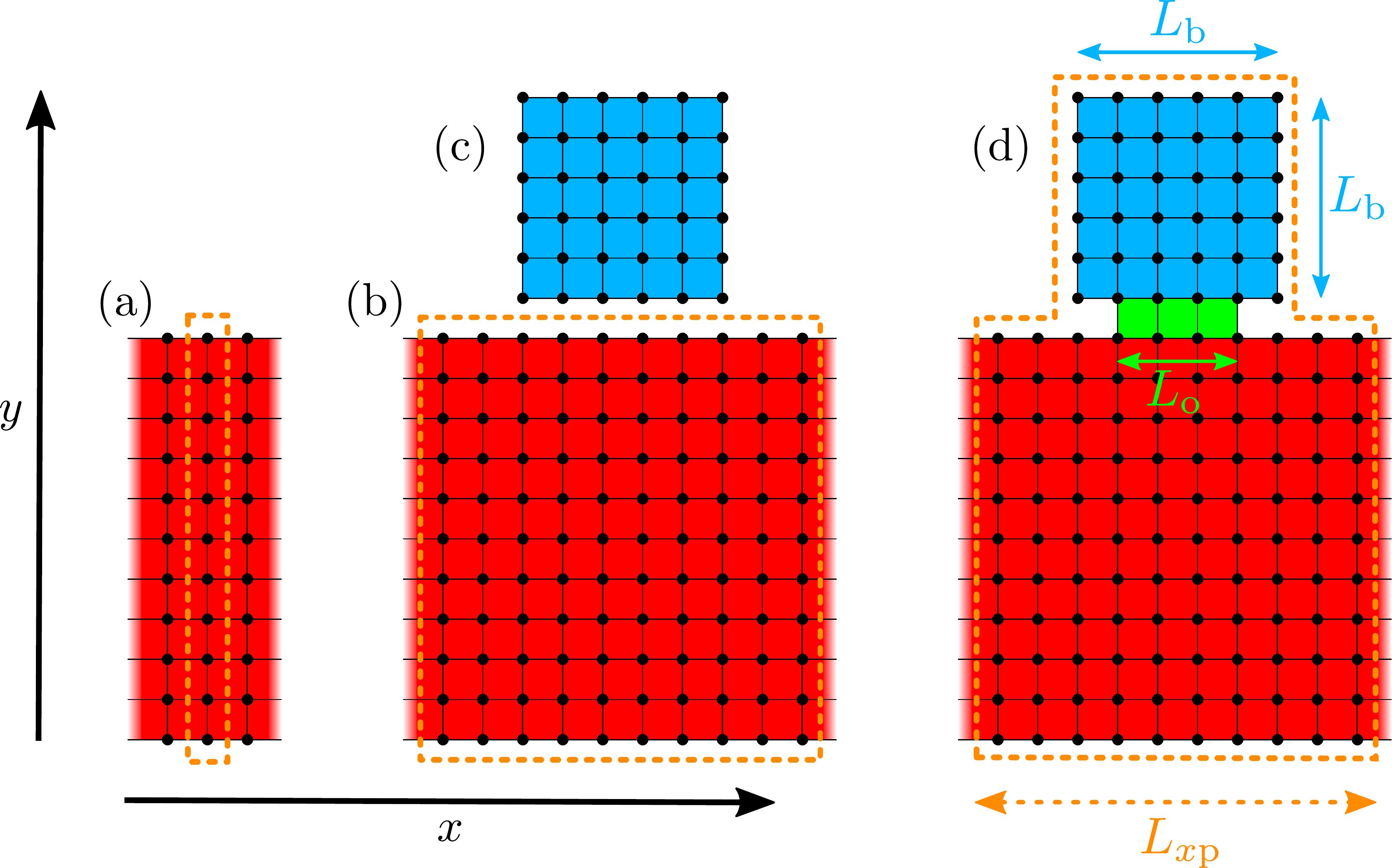} 
	\caption{Sketches of the meshes used to capture the physics
	of the quantum Hall sample with and without bays.  
	The strip geometry (a) and the rectangular geometry (b) are used to describe the 
	sample without bays. The decoupled bay (c) is considered to compute the energy spectrum
	of the isolated bay as reference for the coupled system shown in panel (d).
	The orange dashed lines indicate the respective unit cells.}
	\label{fig:grid}
\end{figure}

In this way, we can very accurately compute eigen energies of the plain
Hall sample as function of $k_x$. In particular, we obtain the wanted
dispersion of the edge states. The level of complexity is illustrated by 
Fig.\ \ref{fig:grid} where the discretization meshes are shown.
The calculation for the plain sample without any decoration only requires
to discretize the $y$-axis, shown in Fig.\ \ref{fig:grid}(a), because the
other spatial dependence is fully captured by the plane wave ansatz \eqref{eq:plane}.
This can be done very efficiently because only a relatively small number 
of sites is required. But in order to be able to later include the bays
as shown in Fig.\ \ref{fig:grid}(d) we first re-calculate the sample without
bays by considering the grid in panel (b).

\subsection{Fully discretized samples}
\label{sec:fully}

Enlarging the unit cell as shown in panel Fig.\ \ref{fig:grid}(b) leads to the 
continuum Hamilton operator 
\begin{equation}
	H_\mathrm{(b)}= \frac{1}{2} \left( \frac{y^2}{\ell_B^4} + 
	2 \mathrm{i} \frac{y}{\ell_B^2} \frac{\partial}{\partial x} - 
	\frac{\partial^2}{\partial x^2}   - \frac{\partial^2}{\partial y^2} \right)
\end{equation}
with the periodic condition for the wave function
\begin{equation}
\label{eq:period}
\psi(x+L_{x\text{p}},y)=\exp(ik_x L_{x\text{p}})\psi(x,y).
\end{equation}
We stress that this condition allows us to determine the value
of $k_x$ only up to multiplies of $2\pi/L_{x\text{p}}$,
as usual if a reduce unit cell in real space is considered.

In the approximate discretized system the first order derivatives are expressed
by using the central finite difference quotient in forth order accuracy
\begin{align}
  \frac{\partial \psi(x, y)}{\partial x}  &\approx \frac{1}{a}
	\left[\frac{1}{12} \psi(x - a2, y) - \frac{2}{3} \psi(x - a, y)
	 + \frac{2}{3}\psi(x + a, y) - \frac{1}{12}\psi(x + 2 a, y) \right] 
\end{align}
wherever possible. Close to a hard-wall boundary the value \smash{$\psi(x + 2 a, y)$} 
is not known because it refers to sites outside of the considered domain. 
Then this term is simply omitted. 
The improvement used for the second derivative based on the
mirroring explained in Fig.\  \ref{fig:spiegel} cannot be used
at hard walls in $x$-direction
because the resulting correction terms would be local densities with
imaginary prefactors spoiling the hermitecity of the Hamiltonian.

Thus, the Hamiltonian $H_\mathrm{(b)}$ is discretized in both directions. 
Expressed in second quantization it is given by
\begin{align}
 H = \sum_{x,y} &\left[ \frac{1}{2} \left( \frac{y^2}{\ell_B^4} + 
\frac{5}{a^2} \right) c_{x, y}^{\dagger} c_{x,y} - \frac{2}{3 a^2} c_{x,y+a}^{\dagger} c_{x,y} + 
\frac{1}{24 a^2} c_{x,y+2a}^{\dagger} c_{x,y} + \frac{\mathrm{i} 2 B y}{3 a} c_{x+a,y}^{\dagger} c_{x,y}\right.  
\nonumber \\
 & \left. - 
\frac{\mathrm{i} B y}{12 a} c_{x+2a,y}^{\dagger} c_{x,y} + \mathrm{h.c.} 
\right]  - \frac{1}{24 a^2} c_{x,b(y)}^\dagger c_{x,b(y)} 
- \frac{1}{24 a^2} c_{b(x),y}^\dagger c_{b(x),y}
 \label{eq:ham_disc}
\end{align}
where $x$ and $y$ run over the discrete sites within the
colored areas in Fig.\ \ref{fig:grid}. The very last term
occurs at hard-wall boundaries in $x$-direction, i.e., 
treating the bays, improving the second derivatives. The periodicity condition
\eqref{eq:period} carries over to 
\begin{equation}
c_{x+L_{x\mathrm{p}}, y} = c_{x, y} \mathrm{e}^{\mathrm{i} k_x L_{x\mathrm{p}}}
\end{equation}
in second quantization.

The Hamiltonian \eqref{eq:ham_disc} can be used to numerically calculate the spectrum for 
any shape of the integer quantum Hall sample. We employ it below to
consider the finite strip without bays first , cf.\ Fig.\ \ref{fig:grid}(b), and
isolated bays, cf.\ Fig.\ \ref{fig:grid}(c), for reference purposes. 
Finally, we pass on to the coupled system, cf.\ Fig.\ \ref{fig:grid}(d).
Then, we also have to include the effect of the gate voltages, see Fig.\ \ref{fig:sample}.
Gate voltage $V_\text{go}$ controls the size of the opening. This is implemented
in our calculation by the choice of the geometry, i.e., by the value of $L_\text{o}$.
Since we only consider bays at the upper boundary there is no $V_{g2}$ to study.
The gate voltage  $V_{g1}$ is implemented by the Hamiltonian part
\begin{equation}
H_\mathrm{bays} = - \sum_{x, y \in\ \mathrm{bays}} 
V_{\mathrm{g}1} c_{x,y}^\dagger c_{x,y}
\end{equation}
where we incorporated the value of the charge into $V_\text{g1}$, i.e., 
we use $V_\text{g1}$ for $|e|V_\text{g1}$.

For small values of $a$ the Hamiltonian \eqref{eq:ham_disc} corresponds to very
large, though sparsely populated matrices. We do not need all eigen values of them
because we focus on the energies of the lowest Landau level up to about the third
Landau level. In particular, the high-lying eigenvalues are strongly influenced
by the discretization and hence they are meaningless for the underlying continuum model.
In order to handle the diagonalization within given intervals of the spectrum 
efficiently for large sparse matrices  we employ the FEAST eigen value solver. 
The FEAST algorithm \cite{poliz09} uses the quantum mechanical density matrix representation 
and counter integration techniques to solve the eigenvalue problem within a given 
search interval. Now, we are in the position to calculate the dispersion of the lowest 
eigen states and thus also able to calculate the Fermi velocities being the
derivatives of the dispersion at the Fermi level.

\section{Dispersions in decorated quantum Hall samples}
\label{sec:dispersion}

So far, we analyzed the Landau levels in the bulk, see Sect.\ \ref{sec:bulk},
and we introduced the approximate Hamiltonians to describe hard-wall boundaries
of varying shapes, see Sects.\ \ref{sec:hardwall} and \ref{sec:fully}. 
Here we present the results for geometries of increasing complexity. First, we address 
the strip geometry, i.e., the sample without any bays. Then, we study the isolated bays
before we address the full coupled system, cf.\ Fig.\ \ref{fig:grid}. 
For clarity, we focus on the lowest Landau levels.

\subsection{Strip geometry}

In the case of a hard-wall confining potential in $y$-direction, i.e., 
$V(y) = \infty$ for $|y| > L_y/2$, one still expects to find eigen values
and eigen states bearing similarities to the bulk solutions. For instance, 
the  eigen function exponentially localized in the middle of the strip
hardly feel the hard-wall confining potential. Hence they closely resemble
the bulk functions \eqref{eq:landau_wave} and their energies are exponentially close
to the bulk Landau levels \eqref{eq:landau_energy}, see also below.

Moreover, the lowest eigen functions localized right at the boundary,
i.e., $k_x = \pm L_2/2 \ell_B^2$, equal the eigen function of the second Landau level $n = 1$.
This is so because the zero of the antisymmetric wave functions coincides with
the boundary \cite{yoshi02} as is well known from the text book problem
in quantum mechanics of a parabolic potential cut off at its apex by an infinite
potential step. Thus, the antisymmetric Hermite polynomials are solutions
which satisfy the boundary condition where they are localized.
The influence of the other boundary is exponentially small if $\ell_B \ll L_y$
which is the limit we presuppose. These special points are used to
verify the accuracy of the calculations based on the discretized model Hamiltonian 
in comparison to the continuum solutions. 

For the discretized description to approximate the continuum efficiently 
in $y$-direction, the distance $a$ between sites must be small enough to capture
the dependence of the Hermite polynomials \eqref{eq:landau_wave} on $y$.
Since $H_n(y)$ has $n$ zeros on the root mean square length $\ell_B \sqrt{n + 1/2}$
we arrive at the constraint
\begin{equation}
\label{eq:y-constraint}
a\ll \ell_B \frac{\sqrt{n + 1/2}}{n+1} \approx \frac{\ell_B}{\sqrt{n+1}}.
\end{equation}
In $x$-direction the wave length set by $2\pi/k_x$ sets an upper limit of $a$
so that we have to require 
\begin{equation}
\label{eq:x-constraint}
a\ll \frac{2\pi}{k_x}.
\end{equation}
While \eqref{eq:y-constraint} needs to be fulfilled in all our calculations,
\eqref{eq:x-constraint} is not required in the solution of \eqref{eq:boundaryterm},
i.e., if the system in Fig.\ \ref{fig:grid}(a) is considered, but only
if the fully discretized model introduced in Sect.\ \ref{sec:fully} is considered.

In addition to these numerical requirements, we argued that we want to
consider the case where the edge states at the upper and at the lower 
boundaries do not interfere. This requires
\begin{equation}
\label{eq:bdry-independence}
\ell_B \frac{\sqrt{n + 1/2}}{n+1} \ll L_y
\end{equation}
on physical grounds. The left hand side is the root mean square of the
spatial extension of the $n$th Landau level in $y$-direction. 
We focus on the lowest bands anyway so that $n=0$ and $n=1$ are the
relevant cases.
For concreteness, we  henceforth use the values $\ell_B=1\mu$m, $a=0.01\ell_B$ and
$L_y=10\ell_B$. These values are in accordance with the above considerations 
for numerical accuracy and independence (up to exponentially small corrections) 
of the edge states.

\begin{figure}[htb]
	\centering
		\includegraphics[width=0.6\columnwidth]{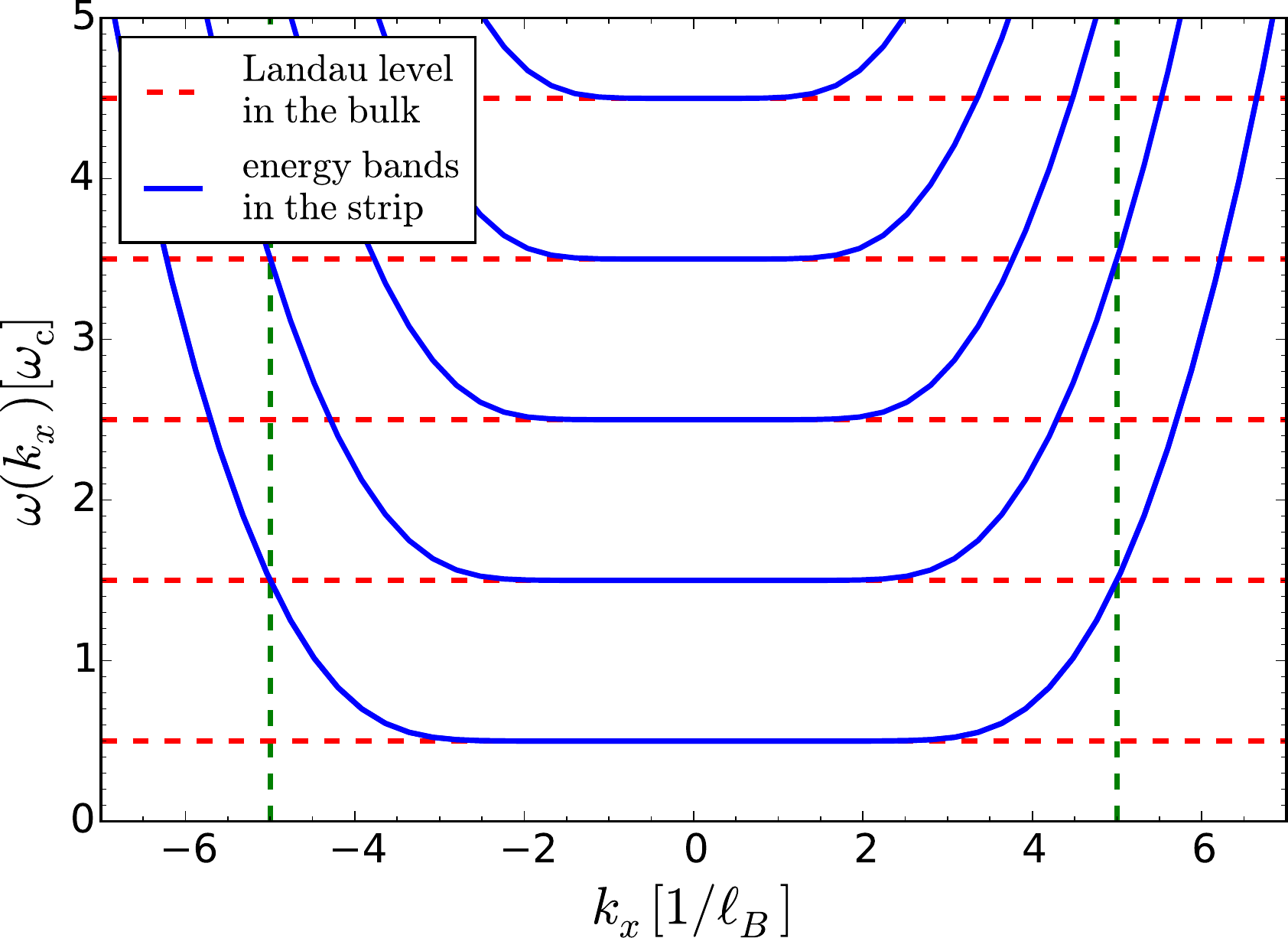} 
	\caption{The blue curves show the dispersions of the Landau levels of a strip of finite \smash{width
	$L_y$}, see Fig.\ \ref{fig:sample}(a). The red dashed lines indicate the equidistant energy spectrum of the Landau levels in the bulk. The vertical dashed lines are located at 
	$k_x = \pm L_y/2 \ell_B^2$ indicating the states which are localized at the upper and lower
	boundaries of the sample.}
	\label{fig:landau}
\end{figure}

Considering the mesh in $y$-direction depicted in Fig.\ \ref{fig:grid}(a) we obtain 
the results (blue solid curves) shown  in Fig.\ \ref{fig:landau} where they
are compared to the bulk results \eqref{eq:landau_energy} (red dashed lines).
Clearly, for small wave vectors one obtains flat bands agreeing very well
with the bulk Landau level. Deviations occur only in the tenth digit of the eigen energies.
This is so because $k_x$ determines the  position
of the harmonic oscillator in $y$-direction, cf.\ Eq.\ \eqref{eq:landau_wave}.
Closer to the boundaries, an upturn in energy occurs
because the electrons feel the hard-wall in their vicinity. As pointed out above,
the state $n=0$ right at the boundary acquires the energy of the Landau level $n=1$
because its wave functions corresponds to half a harmonic oscillator \cite{yoshi02}.
This relation is fulfilled up to the fifth digit thanks to the improved treatment
of the second derivative at the boundary, see Fig.\ \ref{fig:spiegel}.

\begin{figure}[htb]
	\centering
		\includegraphics[width=\columnwidth]{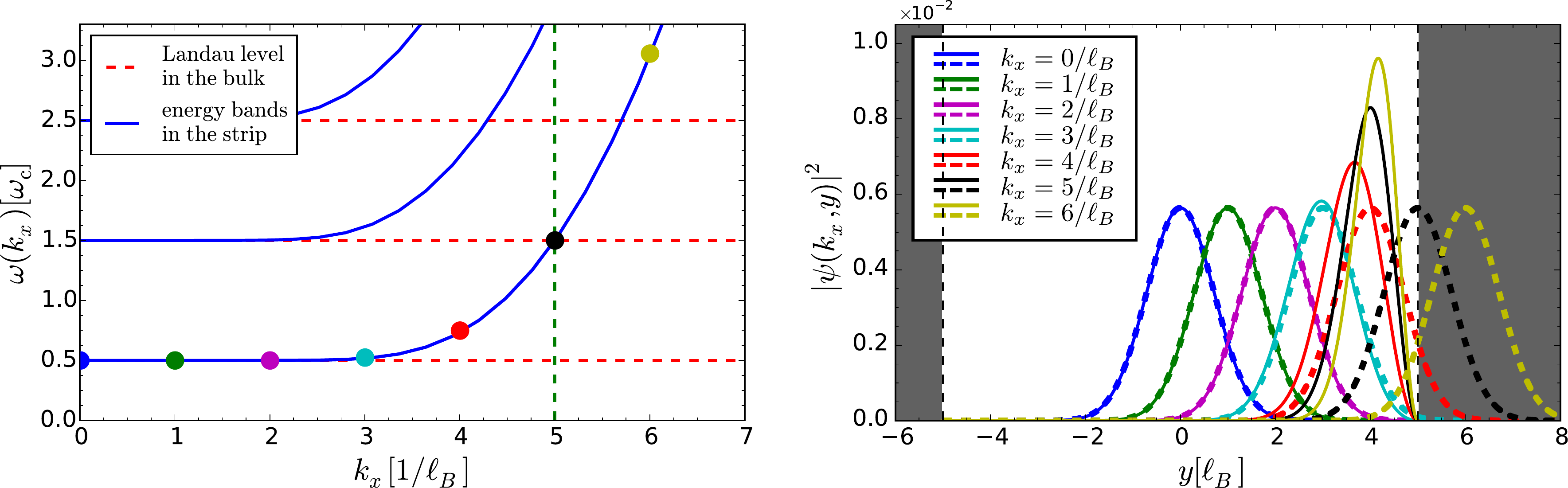}
	\caption{Left panel: Zoom of the dispersion bands in the IQHE.  The vertical dashed green line 
	is located at $k_x = L_y/2 \ell_B^2$ marking the boundary of the sample. The colored dots
	indicate the eigen energies of the corresponding eigen wave functions.
	Right panel:  Probability \smash{densities $|\psi(k_x, y)|^2$} of these eigen wave functions.}
	\label{fig:landau_edge}
\end{figure}

The gradual change of the eigen wave functions upon increasing $k_x$ 
is illustrated in \mbox{Fig.\ \ref{fig:landau_edge}.}
The colored dots in the left panel indicate the energies and the $k_x$ values
of the eigen wave functions depicted in the right panel by solid lines of the
same color. The dashed lines of the same color display the corresponding
eigen functions in the bulk which remain of Gaussian shape throughout.
Note the increase of the peak of the eigen functions in the strip geometry
upon approaching the boundary (sequence red $\to$ black $\to$ yellow) 
because the electron cannot enter the hard-wall.

\subsection{Rectangular geometry}
\label{sec:rectangle}

\begin{figure}[htb]
	\centering
		\includegraphics[width=\columnwidth]{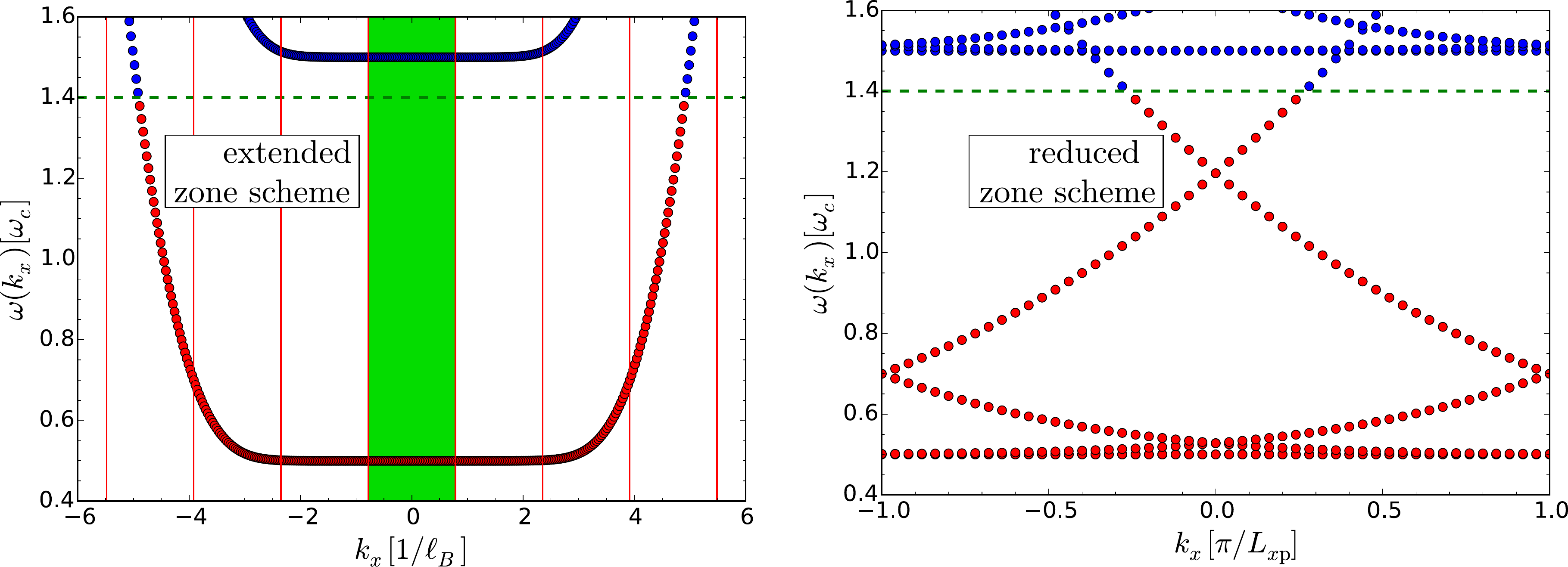} 
	\caption{Left panel: Zoom of the lowest eigen energies in the extended zone scheme 
	for $L_y = \SI{10}{\micro\meter}$ and $L_{x\mathrm{p}} = \SI{4.01}{\micro\meter}$. 
  (the deviation from $\SI{4}{\micro\meter}$ is only due to the discretization).
	Red symbols corresponds to occupied states while blue symbols  represent unoccupied states. 
	The horizontal dashed green line indicates the chosen Fermi energy. 
	The thin vertical red lines show boundaries of the corresponding reduced zone scheme. By backfolding the energies into the green shaded area one obtains the representation of the reduced zone scheme which is shown in the right panel.}
	\label{fig:folded}
\end{figure}

Next, we pass to the fully discretized model \eqref{eq:ham_disc} for the sample without bays,
see Fig.\ \ref{fig:grid}(b). This describes the same physics as the calculation
in the previous subsection. Still, we present exemplary results in Fig.\ \ref{fig:folded}
for two reasons. The first one is to illustrate that this calculation indeed reproduces
the results obtained previously on the mesh Fig.\ \ref{fig:grid}(a)
with sufficient accuracy. Comparing the results from mesh (a) with those from mesh
(b) in Fig.\ \ref{fig:grid} we find that their eigen energies  agree up to the fifth digit.
Note that the calculation for  mesh (a) requires
to deal with a vector space of dimension of the order of 1000 while the calculation for 
\smash{mesh (b)} requires to deal with a vector space with dimension of the order of $10^6$.

The second reason is to obtain results for the undecorated sample, i.e., without bays,
as reference for the subsequent complete analysis. The main point is that the reduction
of the translational invariance by considering the enlarged rectangular unit cell in real space
of \smash{length $L_{x\text{p}}$} leads to a reduced zone scheme in $k_x$ space. The backfolded
branches of the dispersion are shown in the right panel of Fig.\ \ref{fig:folded}. 
Since there is no real, physical reduction of the translational symmetry 
the backfolded branches display level crossings at the boundaries and elsewhere
which are preserved as long as the physical translational symmetry is preserved.
Hence the backfolded branches can be unfolded again to yield the extended zone scheme
display in the left panel of Fig.\ \ref{fig:folded}. This shows the same results as 
were obtained directly by the previous calculation based on mesh Fig.\ \ref{fig:grid}(a),
presented in Figs.\ \ref{fig:landau} and \ref{fig:landau_edge}.

For clarity, we have chosen in Fig.\ \ref{fig:folded} to consider a quantum
Hall sample of finite length. The length of the unit cell in real space
is given by $L_{x\mathrm{p}}$ and we fix the total number of these cells
to $N_x=50$. Of course, this value can easily be changed if needed.
Hence, there are $N_x$ different momenta $k_x$
in the reduced zone scheme. They are multiples of $2\pi/N_xL_{x\mathrm{p}}$
lying in the interval $\left[ -\pi/L_{x\mathrm{p}}, \pi/L_{x\mathrm{p}} \right]$.

We want to focus on the filled lowest Landau level, i.e., filling factor $\nu=1$. Due to 
the upturn of the lowest level upon approaching the boundaries of the sample this 
filling factor requires to occupy all states with energies just below the flat region
of the second lowest level, see left panel of Fig.\ \ref{fig:folded}. 
However, in order to 
exclude any spurious effects of the energy levels of the second lowest Landau level
we set the Fermi level to a value slightly below the flat band of the Landau
level $n=1$, namely to $\epsilon_\text{F} = 1.4 \omega_{\mathrm{c}}$ as indicated 
by the green dashed line in \mbox{Fig.\ \ref{fig:folded}.} This allows us to distinguish
unambiguously between occupied and unoccupied levels.
This procedure helps to identify our quantity of interest, the Fermi velocity, 
i.e., the derivative of the dispersion with respect to $k_x$ at the Fermi level.
 The ensuing minor deviation of
the filling factor $\nu$ from 1 is macroscopically irrelevant for large values of $L_y$.

\subsection{Isolated bays}

Before dealing with the complete system with bays coupled to the 
quantum Hall sample we determine the energy spectrum of isolated bays
for later comparison. Note that we choose to consider quadratic bay for
calculational simplicity. But the underlying physics does not require a particular
shape of the bay so that samples decorated with circular bays
will show the same physics at somehow modified quantitative parameters.

\begin{figure}[htb]
	\centering
		\includegraphics[width=0.5\columnwidth]{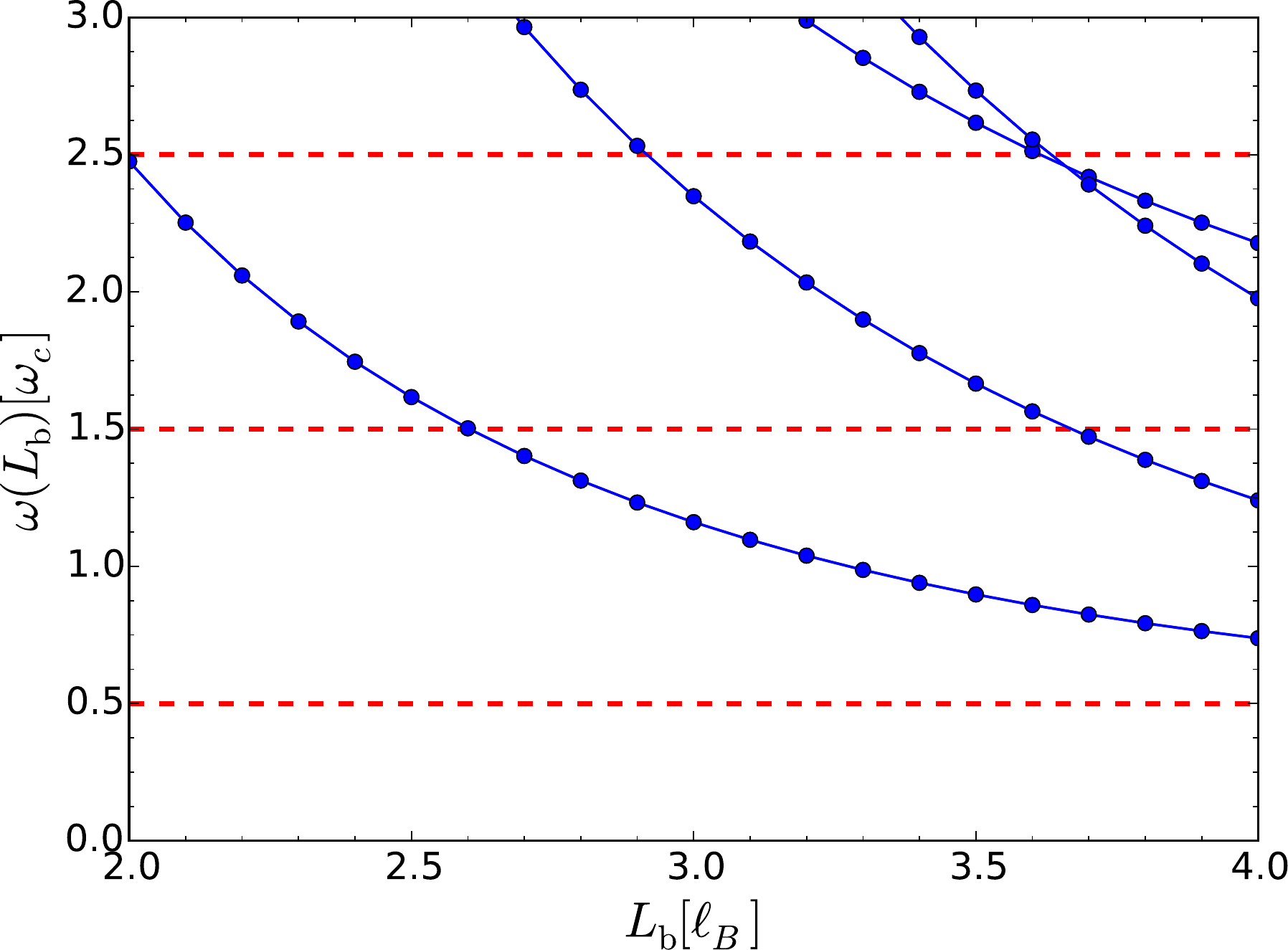} 
	\caption{Discrete energy spectra of decoupled, i.e., isolated  bays 
	as function of their size $L_\mathrm{b}$ are rendered by blue solid lines and blue
	symbols for $\ell_B = \SI{1}{\micro\meter}$. The horizontal red dashed lines indicate
	the equidistant Landau levels in the bulk for comparison.}
	\label{fig:bay}
\end{figure}

For considering the isolated bays we treat the mesh shown in Fig.\ \ref{fig:grid} (c). The calculated energy spectrum as function of the  length $L_\mathrm{b}$ is plotted in Fig.\ 
\ref{fig:bay}. Having the classical cyclotron picture of circular electronic orbits in mind 
we choose $L_\mathrm{b} = 2 \ell_B$ as starting value. No smaller bay would allow
for a classical circular orbit. As expected the energies are larger than the
bulk Landau energies because the confinement due to the bays restricts the
motion of the electrons. Accordingly, increasing $L_\mathrm{b}$ lowers the energies because
enlarging the bays reduces the influence of the confining potential. 

 The lowest eigen energy of the bay reaches the energy gap between the two lowest Landau levels 
at a bay size of $L_\mathrm{b} \approx 2.6 \ell_B$. Using the gate voltage $V_\text{g1}$ to 
shift the energies in the bays relative to the rest of the sample offers a
possibility to tune a local mode in resonance to an edge mode. We will discuss
this in more detail in the next subsection.

Adding the decoupled bay to the unit cell, i.e., considering the model
shown in the \smash{panels (b)} and (c) of Fig.\ \ref{fig:grid} without any coupling
yields the eigen energies provided in \mbox{Sect.\ \ref{sec:rectangle}} plus the eigen energies
of the bays which do not disperse at all (not shown). They appear as completely flat
modes if plotted against $k_x$ due to their completely local nature in real space.

\subsection{Quantum Hall sample with coupled bays}

Now, we pass to the fully decorated sample where the bays are coupled
to the 2D electron gas in the strip, i.e., we consider the mesh in Fig.\ \ref{fig:grid}(d).
We switch on the coupling between the bays and the strip by gradually
increasing the opening $L_\mathrm{o}$ from zero to the maximum \smash{value $L_\text{b}$.}
The energy spectra are computed and tracked to understand how the coupling
influences the eigen states in general and the edge modes in particular.

\begin{figure}[htb]
	\centering
		\includegraphics[width=\columnwidth]{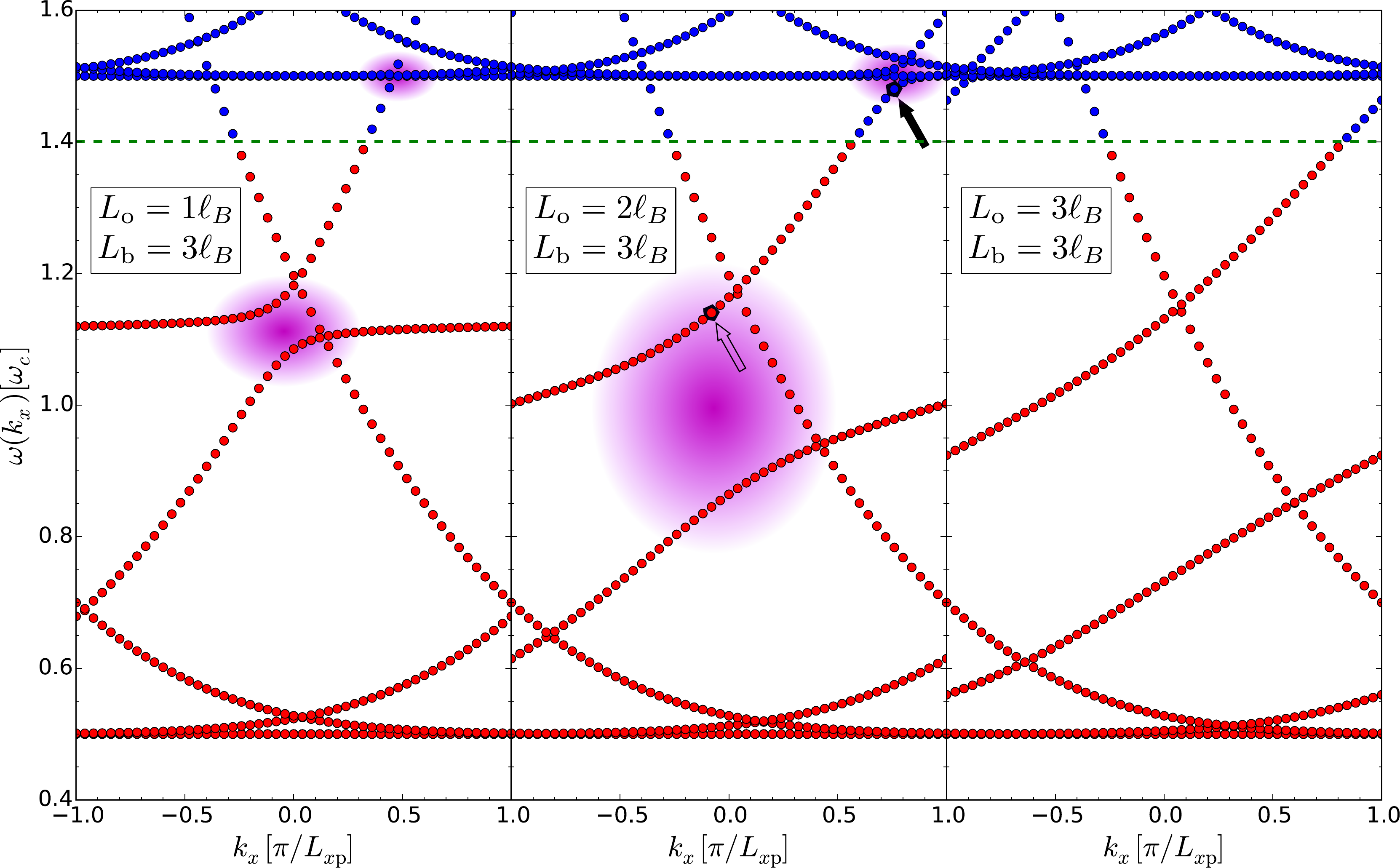} 
	\caption{Energy spectra of the lowest eigen states of a quantum Hall sample with 
	$L_y =  \SI{10}{\micro\meter}$, $L_{x\mathrm{p}} = \SI{4.01}{\micro\meter}$, and 
	$\ell_B = \SI{1}{\micro\meter}$. The left panel shows the case of weakly coupled bays
	because the opening $L_\text{o}$ is small. The middle panel shows a moderate coupling
	while the right panel a rather strongly coupled case because the opening $L_\text{o}$
	is increased step by step. Red symbols correspond to occupied states while blue symbols
	depict unoccupied states; the dashed horizontal green line indicates the chosen Fermi level.
	The shaded areas highlight the locations of two avoid crossings due to the hybridization
	of local and dispersive modes.}
	\label{fig:coupled}
\end{figure}

To this end, we depict three representative cases with openings 
$L_\mathrm{o} = \left\lbrace 1 \ell_B, 2 \ell_B, 3 \ell_B \right\rbrace$ 
and a bay size $L_\mathrm{b} = 3 \ell_B$ in Fig.\ \ref{fig:coupled}. 
They represent the cases of weak, moderate, and strong coupling of the bays.
Upon coupling the bays to the quantum Hall sample, i.e., for $L_\mathrm{o} \neq 0$, 
the eigen states of the bays and the strip start to merge.
Energy crossings of local modes from the bays with dispersive edge modes
in absence of any coupling turn into avoided crossings once
the bays and the strip are coupled.
This represents a clear finger print of level repulsion. 

Inspecting the three panels, one realizes that only the right moving
edge modes are influenced by the coupling of the bays. Only their energies
depend on the degree of coupling, i.e., on the size $L_\text{o}$ of the opening.
The left moving modes are spatially separated because they are localized
at the other boundary of the sample without decoration. Hence they are 
influenced only exponentially weakly.

\begin{figure}[htb]
	\centering
		\includegraphics[width=0.75\columnwidth]{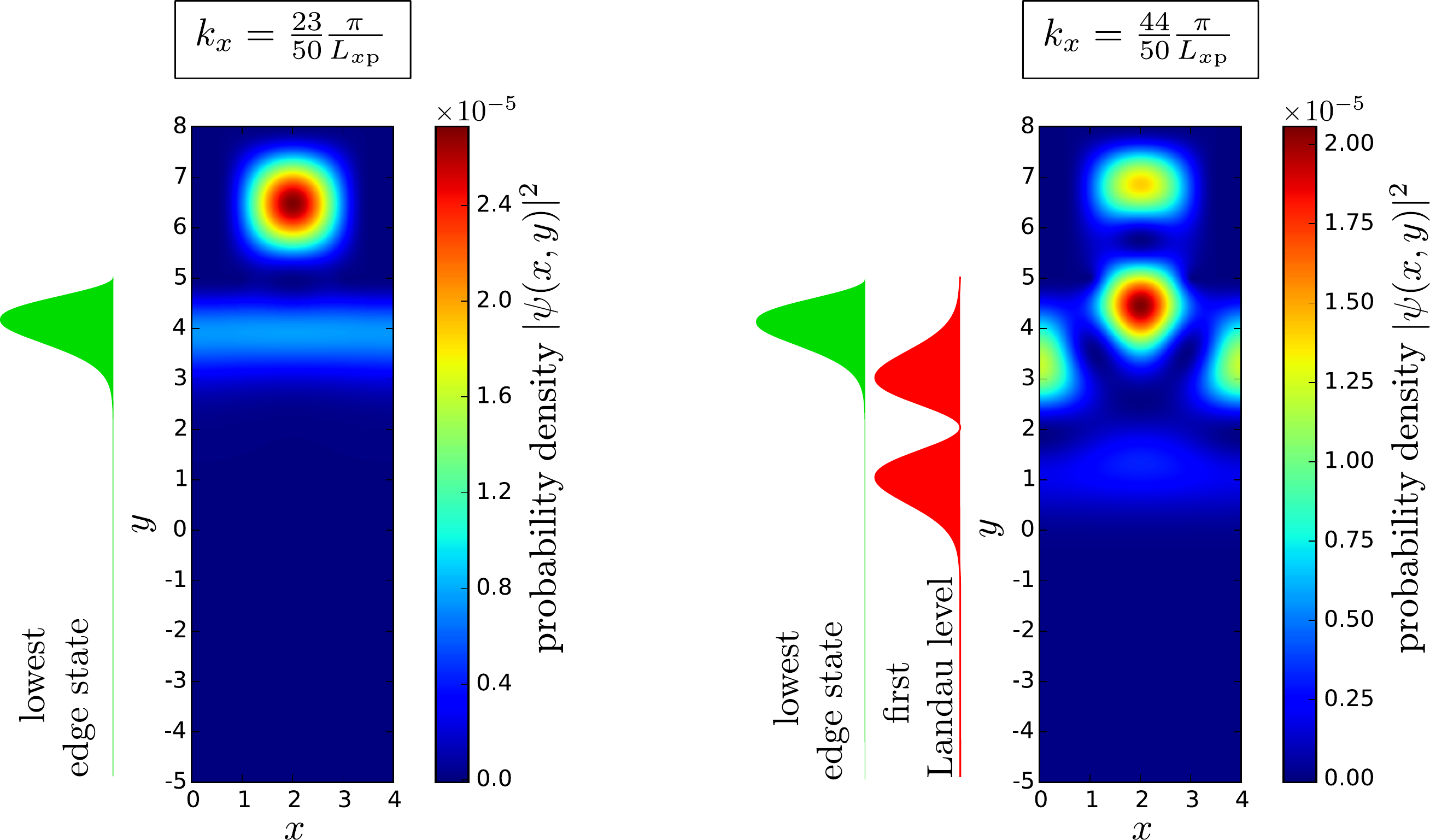} 
	\caption{Probability density $|\psi_n(x, y)|^2$ of two eigen states influenced by 
  the different avoid crossings. The left panel shows the hybridization between the 
	edge state of the Landau level $n=0$ with the local mode in the bay. The energy and
	momentum of this state are indicated in the middle panel of Fig.\ \ref{fig:coupled}
	by the open arrow.
	The right panel shows the weak hybridization of the edge state with the second Landau level
	$n=1$ mediated by the local mode in the bay. 
	The energy and 	momentum of this state are indicated in the middle panel of 
	Fig.\ \ref{fig:coupled} by the filled arrow.
	The parameters of the geometry are $L_y = \SI{10}{\micro\meter}$, $L_{x\mathrm{p}} = 
	\SI{4.01}{\micro\meter}$, $L_\mathrm{b} = \SI{3}{\micro\meter}$, $L_\mathrm{o} = 
	\SI{2}{\micro\meter}$, and $\ell_B = \SI{1}{\micro\meter}$.}
	\label{fig:psi}
\end{figure}

A nice example of the level repulsion between a (formerly) local bay mode
and a dispersive, right moving edge mode is seen in the middle of the panels in Fig.\
\ref{fig:coupled} around $k_x=0$. The relevant area is shaded in violet in the 
left and the middle panel. An example of a corresponding
wave function is shown in the left panel of Fig.\ \ref{fig:psi}.
 In the right panel of \mbox{Fig.\ \ref{fig:coupled}} the avoided crossing is 
still present, but hardly discernible because the energies are 
already very different due to the strong coupling.
In return, the left panel shows the character of an avoided level 
crossing most clearly because the coupling of the bays is still small
and hence the hybridization between the bay modes and the strip modes is still
small.

Another, less obvious and thus surprising, origin of avoided level crossings 
between dispersive edge modes and local modes results from the breaking of the translational
invariance and the concomitant backfolding. This mechanism induces hybridization
between local Landau levels and edge modes. An example is indicated by
a shaded area in the left panel at $k_x\approx0.4 \pi/L_{x\text{p}}$ and 
in the middle panel at $k_x\approx0.8 \pi/L_{x\text{p}}$ of Fig.\ \ref{fig:coupled}. 
Clearly, the effect is weaker
than the hybridization of edge modes and local bay modes. 
This is so because the coupling of edge modes and local Landau levels
is a second order effect in the coupling of the bays to the strip.
The bay modes are involved only indirectly by virtual processes, see also
the right panel of Fig.\ \ref{fig:psi} where an exemplary wave function
is shown.
 Similar effects were also found in the IQHE where different edge modes 
start to mix with one other due to breaking the translational symmetry 
by a step potential \cite{ventu11}.

To support the interpretations given above, we plot the probability density 
$|\psi(x, y)|^2$ for eigen states from the two avoided level crossings in 
Fig.\ \ref{fig:psi}. The left panel shows a state built from an edge mode 
and a local mode from the bays; its position in the energy spectrum
is indicated by a solid arrow in the middle panel of Fig.\ \ref{fig:folded}.
Clearly, the two constituents, the edge mode and the local mode in the bay
can be seen.

The right panel Fig.\ \ref{fig:psi} shows a state built from an edge mode, 
a local mode from the bays, and a the next higher Landau level $n=1$; 
its position in the energy spectrum
is indicated by a filled arrow in the middle panel of \mbox{Fig.\ \ref{fig:folded}}.
Here, three states are involved and contribute to the eigen states
as can be discerned nicely. The contribution of the local mode in the bay
is much smaller than in the case shown in the left panel because it
contributes only as virtual state mediating the breaking of the translational
invariance.

\section{Tuning the Fermi velocity}
\label{sec:tune}

In the previous sections we developed a detailed understanding
of the energy spectra of quantum Hall sample decorated by bays.
Our ultimate goal is to study whether and how the Fermi velocity
$v_\text{F}$
can be tuned in such a decorated quantum Hall sample. 
We highlight that the Fermi velocity $v_\text{F}$ 
represents the group velocity of the coherent quantum mechanical
propagation of electronic wave packets. It cannot be seen as
classical propagation of electrons along the (longer) boundaries
of the bays, see below.
Here we present quantitative results for the Fermi velocity and its
dependence on the parameters of the model.

\begin{figure}[htb]
	\centering
		\includegraphics[width=\columnwidth]{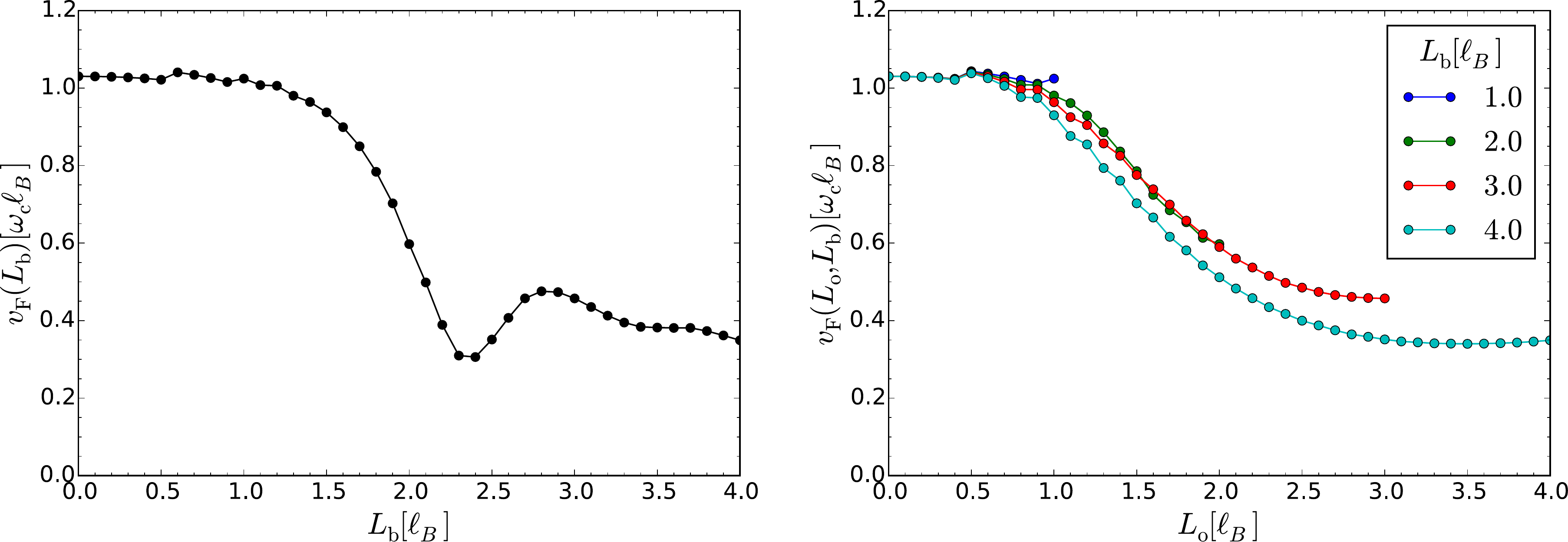} 
	\caption{Left panel: Fermi velocity $v_\mathrm{F}$ of the right moving modes
	as function of the bay size $L_\mathrm{b}$ with 
	at $L_y = \SI{10}{\micro\meter}$, $L_{x\mathrm{p}} = \SI{4.01}{\micro\meter}$, 
	$\ell_B = \SI{1}{\micro\meter}$, and $L_\mathrm{o} = L_\mathrm{b}$. Right panel: Fermi \smash{velocity $v_\mathrm{F}$} as a function of the opening $L_\mathrm{o}$ of the bays
	for various bay \smash{sizes $L_\mathrm{b}$} with \smash{$L_y = \SI{10}{\micro\meter}$,} $L_{x\mathrm{p}} = 
	\SI{4.01}{\micro\meter}$, and $\ell_B = \SI{1}{\micro\meter}$.}
	\label{fig:vf_bay_Lo}
\end{figure}

First, we examine the dependence of $v_\text{F}$ on the size of the bays by 
increasing $L_\mathrm{b}$ for maximally opened bays, i.e., for $L_\mathrm{o} = L_\mathrm{b}$.
The results are shown in the left panel of \mbox{Fig.\ \ref{fig:vf_bay_Lo}.}
For maximally opened bays $L_\mathrm{o} = L_\mathrm{b}$ the dispersions display no 
flat region because the strong level repulsion induces 
sizable momentum dependencies of most modes,
see right panel of  Fig.\ \ref{fig:coupled}. Thus no strong dependence of the
Fermi velocity is expected in accordance with the left panel of Fig.\ \ref{fig:vf_bay_Lo}.
The complex interplay of many hybridizing levels makes it impossible to  predict sizes 
for which parameter precisely $v_\mathrm{F}$ takes its minimum value.
However, the comparison of the left panel in Fig.\ \ref{fig:vf_bay_Lo} with  Fig.\ \ref{fig:bay}
reveals that the Fermi velocity is indeed influenced when the local mode
in the bay approaches the Fermi level, here $1.4\omega_\text{c}$, which is the case
around $L_\mathrm{b} = 2.6 \ell_B$. Note that the Fermi velocity
is generally reduced, roughly by a factor 2, once the local modes have come down 
in energy so that they reach the Fermi level.

The next parameter varied is the opening $L_\text{o}$ of the bay.
The right panel of Fig.\ \ref{fig:vf_bay_Lo} shows the results for various bay sizes. Note that the opening 
cannot exceed the size of the bay, hence the curves stop at $L_\text{o}=L_\text{b}$.
All curves follow the general trend that the Fermi velocity is lowered upon
increasing the hybridization between local modes in the bays and the dispersive
edge modes. This is achieved by increasing the opening $L_\text{o}$.
An approximate reduction by a factor of 2 is achieved once the
local energy levels from the bay come down in energy, i.e., for large \smash{enough
$L_\text{b}$.} This reduction is not very impressive; in addition, the geometry 
is fixed once the sample is grown and cannot be tuned on the fly.

\begin{figure}[htb]
	\centering
		\includegraphics[width=0.5\columnwidth]{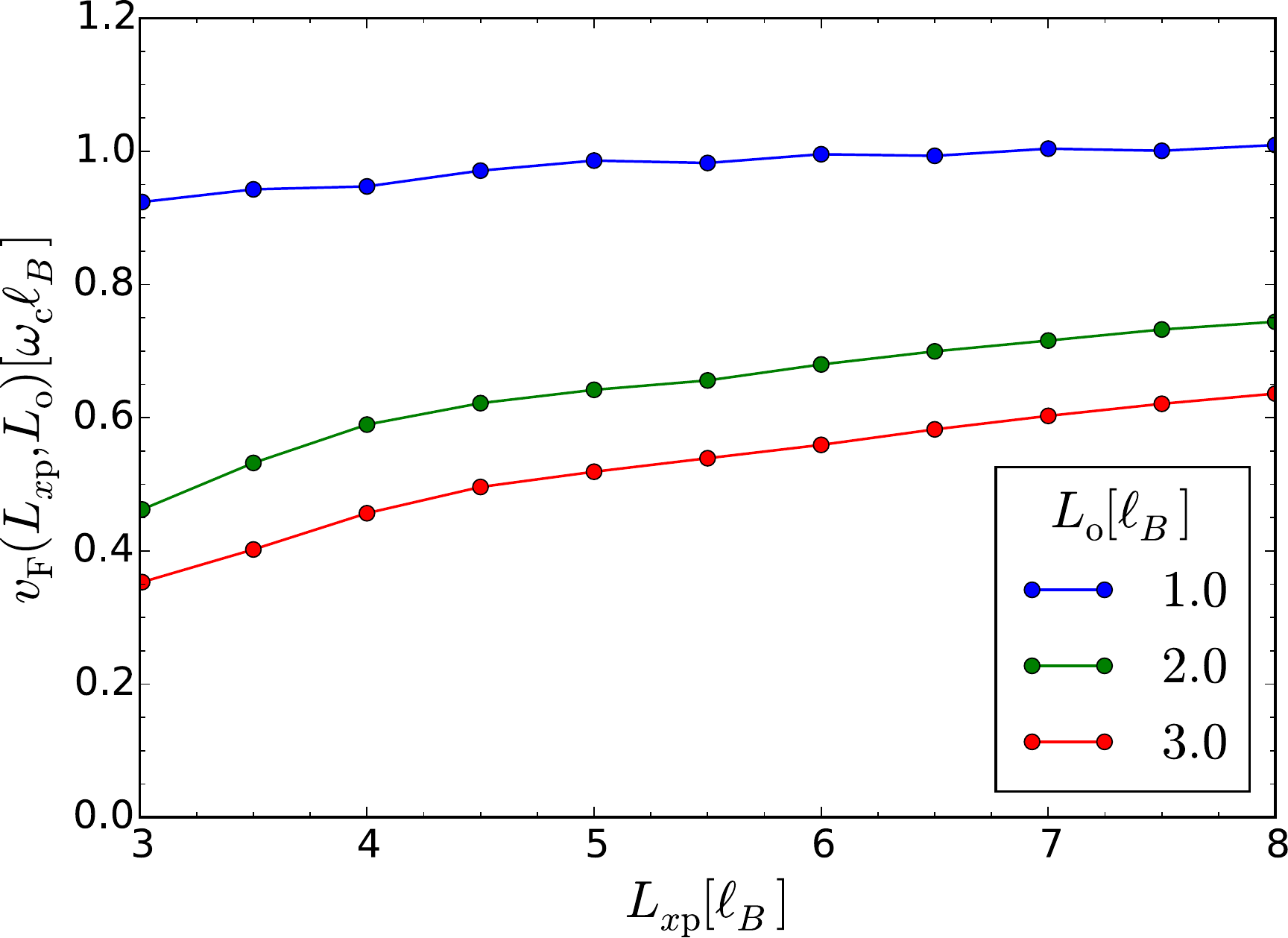} 
	\caption{Fermi velocity $v_\mathrm{F}$ as function of the distance between the bays, i.e.,
	$L_{x \mathrm{p}}$, see \mbox{Fig.\ \ref{fig:ez},} for various bay openings $L_\mathrm{o}$ with 
	$L_y = \SI{10}{\micro\meter}$, $L_{\mathrm{b}} = \SI{3}{\micro\meter}$, and 
	$\ell_B = \SI{1}{\micro\meter}$.}
	\label{fig:vf_Lxp}
\end{figure}

The last dependence of $v_\text{F}$ on a geometric parameter, that we study, is the dependence
on the distance between the bays, i.e., the size $L_{x\text{p}}$, of the decorated unit cell,
see Fig.\ \ref{fig:ez}. One could imagine that a certain resonance phenomenon occurs
for special values of $L_{x\text{p}}$. Generally, we expect that the influence of the 
decorating bays decreases upon increasing $L_{x\text{p}}$ because the fraction 
of decorated boundary decreases. Explicit results are shown in Fig.\ \ref{fig:vf_Lxp}.
Again, the dependence of $v_\text{F}$ is rather weak. The expected trend 
that larger $L_{x\text{p}}$ reduces $v_\text{F}$ less 
is clearly confirmed because the Fermi velocity approaches
its undecorated value of \smash{about $1\omega_\text{c}\ell_B$} upon increasing $L_{x\text{p}}$.
At small values of $L_{x\text{p}}$ we retrieve a reduction of the order of a
factor 2. But no resonance phenomena at particular values of the interbay distance
are found. We attribute this to the fact that none of the local modes in the bay
is truly in resonance with the edge modes

In order to identify a suitable tuning parameter we resort to the results gained
for lattice models \cite{uhrig16,malki17b}. Three ingredients are important for
sizable changes of the Fermi velocity: (i) the local and the dispersive modes
must be in (or close to) resonance. (ii) There must be a parameter to tune and to
detune this resonance. (iii) The coupling of the modes should be rather small
so that they are sensitive to being or not being in resonance.

Translating these conclusions back to the IQHE, it appears that we have to use the
gate voltage $V_\text{g1}$ to control the resonance between the local modes
in the bays and the dispersive edge modes. It is obvious that one can shift
the bay modes by changing $V_\text{g1}$. An additional asset is that this can be
done on the fly so that one disposes of a true control knob for the speed of
signal transmission and hence for the delay time which can be turned while
the signal processing is going on.

\begin{figure}[htb]
	\centering
		\includegraphics[width=0.5\columnwidth]{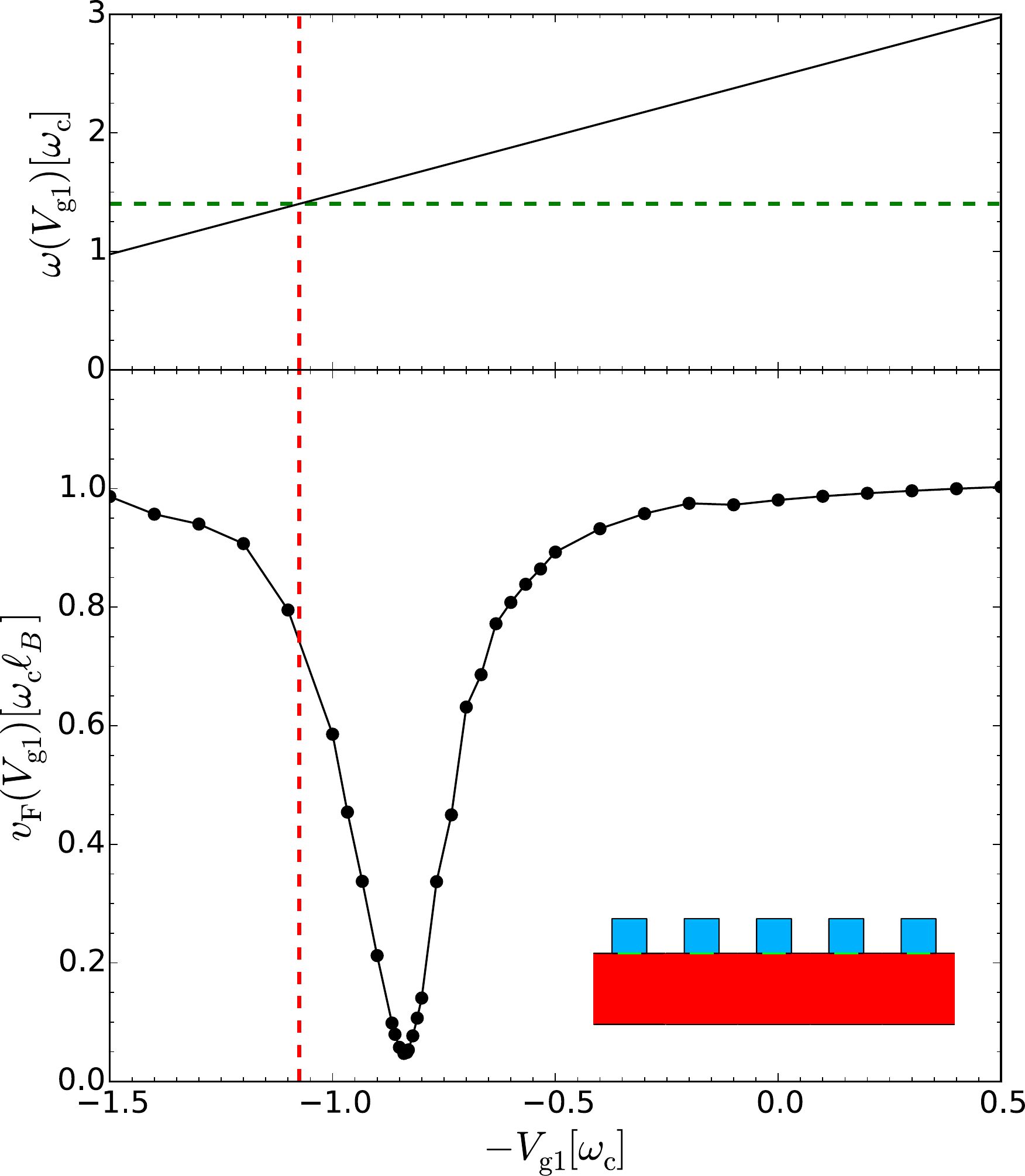} 
	\caption{Upper panel: The horizontal dashed line shows the set Fermi level while
	the slanted solid lines depict the energy level in the isolated bays shifted by
	the gate voltage. The vertical red line is a guide to the eye to link the resonance
	visible in the upper panel to the strong response in the lower panel.
		Lower panel: Fermi velocity $v_\mathrm{F}$ as function of the gate 
		\smash{voltage $V_\text{g1}$}
	for $L_\text{b}=\SI{2}{\micro\meter}$, 
	$L_y = \SI{10}{\micro\meter}$, $L_{x\mathrm{p}} = \SI{4.01}{\micro\meter}$, 
		$L_\mathrm{o} = \SI{1}{\micro\meter}$, and $\ell_B = \SI{1}{\micro\meter}$.}
	\label{fig:vf_pot1}
\end{figure}

The opening of the bays should not be large because the coupling and hence
the hybridization of the local and the dispersive modes should be rather weak.
Thus we choose  the rather small value $L_\text{o}=\ell_B$ in Fig.\ \ref{fig:vf_pot1}. 
In this figure, we plot the dependence of the Fermi velocity on
the gate voltage. For most values, the Fermi velocity does not deviate strongly from
its value of about $1\omega_\text{c}\ell_B$ in a sample without bays. But if the energy levels of
the local modes in the bays approach the dispersive edge mode at the Fermi level
they resonate and produce an avoided level crossing. In the region of the
 avoided level crossing
the local mode and the dispersive one mix so that the formerly steep
crossing of the dispersion through the Fermi level becomes flat. Hence the Fermi velocity 
is considerably suppressed. Note that the resulting resonance dips of
$v_\text{F}$ are rather narrow and can easily be used to (de)tune the velocity
by moderate changes of the applied external gate voltage.
In this fashion, changes of the Fermi velocity by factors 10 to 100 should
be realizable, similar to what was found in lattice models \cite{uhrig16,malki17b}.

\begin{figure}[h!]
	\centering
		\includegraphics[width=0.8\columnwidth,clip]{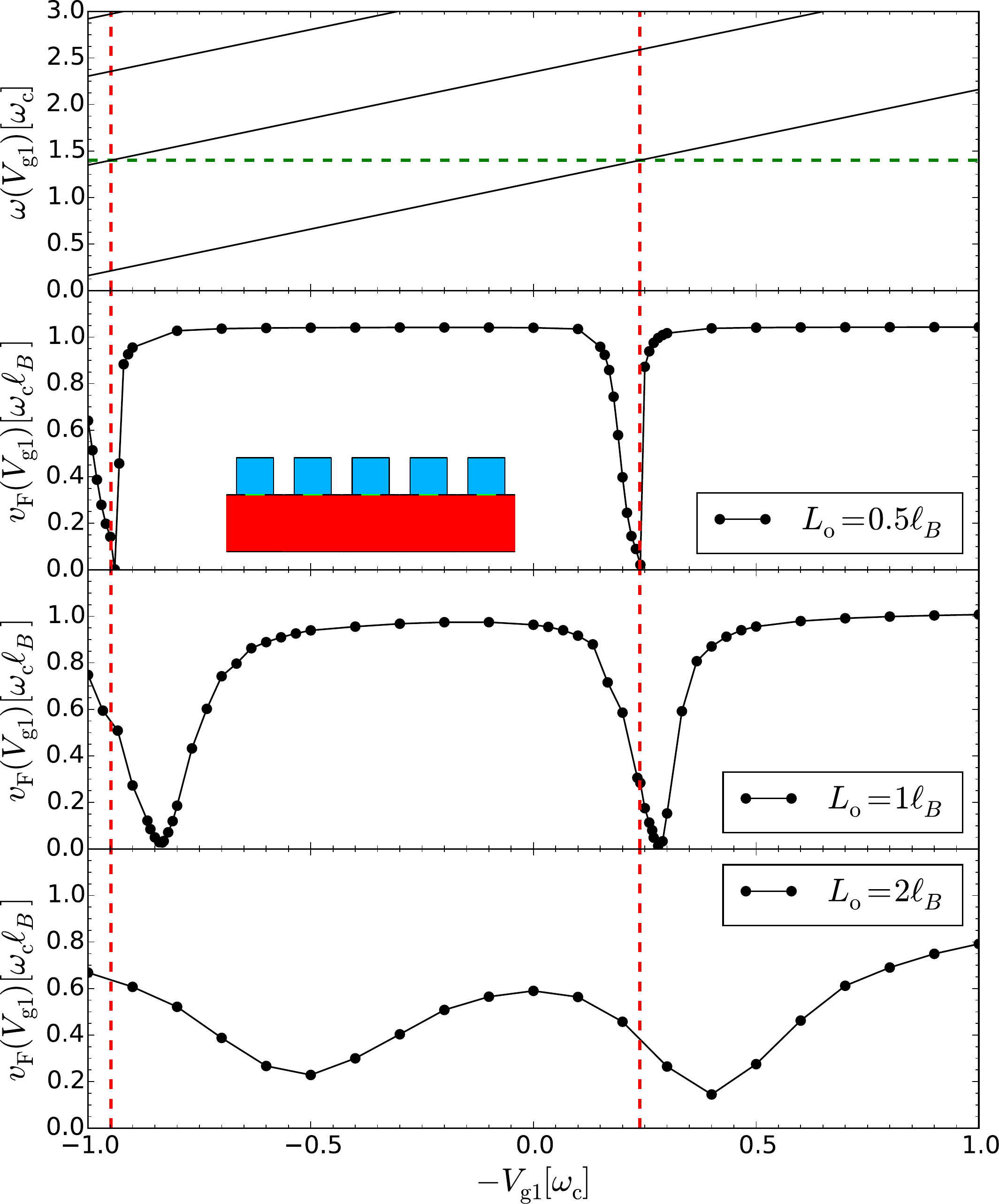} 
	\caption{Upper panel: The horizontal dashed line shows the set Fermi level while
	the slanted solid lines depict the energy level in the isolated bays shifted by
	the gate voltage. The vertical red lines are guides to the eye to link the resonance
	visible in the upper panel to the strong response in the lower three panels.
		Lower three panels: Fermi velocity $v_\mathrm{F}$ as function of the gate 
		\smash{voltage $V_\text{g1}$} for $L_\text{b}=\SI{3}{\micro\meter}$, 
	$L_y = \SI{10}{\micro\meter}$, $L_{x\mathrm{p}} = \SI{4.01}{\micro\meter}$, 
	 $\ell_B = \SI{1}{\micro\meter}$, and the three different values of	
	$L_\mathrm{o} = \SI{1}{\micro\meter}$ as indicated.	
	\label{fig:vf_pot2}}
\end{figure}

Comparing Figs.\ \ref{fig:vf_pot1} and \ref{fig:vf_pot2} one realizes the
similarities of the curves. The width of the resonance dips is comparable 
if the openings $L_\text{o}$ are the same, cf.\  Fig.\ \ref{fig:vf_pot1}
and the second lowest panel of Fig.\ \ref{fig:vf_pot2}.

Fig.\ \ref{fig:vf_pot2} illustrates very clearly, that 
larger openings lead to significantly broader dips which are less
deep. In return, smaller openings and thus less coupled bays lead
to narrower dips with significantly lower residual Fermi velocity 
at the minimum. This minimum value of $v_\mathrm{F}$ 
depends on how flat the dispersion of the hybridized modes remains
as determined by the coupling strength: Weaker coupling implies 
better localized hybridized modes with flatter dispersion. 
Flatter modes allow for sharper dips to lower residual values of the
Fermi velocity. Note that the reduction of the Fermi velocity can reach a factor of
100 for narrow bay openings. A classical interpretation of slower
propagation due to the longer path along the boundaries of the
bays would explain a factor $3.75$ at best for $L_\text{o}=
\SI{0.5}{\micro\meter}$.

The positions of the resonance dips
depend on the energy levels of the modes in the bay so that
the bay size influences them strongly. In the smaller bays studied
in Fig.\ \ref{fig:vf_pot1}, the lowest bay level lies above the Fermi
level so that the gate voltage has to bring it down in order to observe
resonance.  In the larger bays, studied
in Fig.\ \ref{fig:vf_pot2}, the lowest bay level lies below the Fermi
level while the second lowest above it.
So Fig.\ \ref{fig:vf_pot2} shows that several dips may occur, even
for different signs of the gate voltages. 
All in all, it appears that the precise position of the dips is not
at the resonance of the energy levels of the decoupled bays, but at slightly
higher values of the gate voltage. We attribute this to the effects
of the hybridizing couplings which shifts the local modes in the bays
downwards in energy.

\section{Conclusions}
\label{sec:conclusion}

Topologically protected edge states possess many theoretically appealing properties.
Still, avenues towards applications have not been followed by broad research.
The recent proposal of tunable Fermi velocities in Chern insulators and
spinful topological insulators for the realization of delay lines and interference
devices is a step in this direction. The purpose of the present study
was to show that no lattice models are required, but that semiconductor samples
with decorated boundaries show the same phenomena. This finding represents 
a substantial step forward towards realization because of the extremely 
high standard of designing and growing nanostructures for semiconductor devices.

We analyzed the dependence of the dispersion of the edge states in decorated 
quantum Hall samples on various
parameters. The geometry of the sample sets the energy levels and partly
the degree of coupling between the decorating bays and the bulk of the
two-dimensional sample. Yet the geometric parameters do not allow for
a fine-tuning of the Fermi velocity, let alone quick changes of it in the course
of signal processing. 

But gate voltages can achieve the wanted tunability. First, we found that the
local levels in the bays should be close in energy to the Fermi level in
the remainder of the quantum Hall sample so that the gate voltage applied to
the bays does not need to shift them to a large extent. Second, the coupling
between the bays and the rest of the sample should be rather weak to 
have rather narrow and deep dips in the Fermi velocity if the local
modes are tuned into resonance to the dispersive edge states.
Then, the fundamental mechanism of mode mixing and level repulsion
leads to weakly dispersive eigen modes crossing the Fermi level.
This represents the key phenomenon for tunability.

Changes by up to two orders of magnitude appear possible.
In our calculations, the degree of coupling is a geometric parameter.
In practice, we propose to make it tunable as well by additional gate
electrodes which modify the width of the opening of the bays, cf.\
Ref.\ \cite{feve07}. 

The calculations are based on discretizing the sample in real space
and mapping it to a tight-binding type of model. For fine enough meshes, reliable
results valid for the continuum case are obtained
as we could verify by comparison to analytic bulk solutions. 
We increased the complexity of the considered geometry step by step in order
to gain a reliable understanding of the occurring physical phenomena.
The approach is flexible enough to be adapted to various geometries. 
We considered quadratic bays, but any other shape is possible as well,
but only small, quantitative changes are expected.
Here the focus was
on a proof-of-principle calculation to show that the anticipated physics 
takes indeed place in the integer quantum Hall effect.

In view of experimental realizability, some aspects must be kept in
mind. First, the neglected interaction between the electrons may lead to
the formation of certain charge modulations at the boundary. On the one hand, 
it is established that compressible and incompressible stripes form close
to the boundaries \cite{chklo92}. The incompressible stripes may
hinder the propagation of signals. On the other hand, if the filling 
is tuned just below filling factor $\nu=1$, we expect that this
effect is avoided because no incompressible stripes are formed at the
edges. The final clarification, however, can only be reached by
an experimental study.

For concreteness, we showed calculations for 
$\ell_B=\SI{1}{\micro\meter}$. This 
value corresponds via $B=\hbar/(e\ell_B^2)$ to a magnetic field of
$0.66$mT and to a electron density of $3.2\cdot 10^7$cm$^{-2}$. 
Both values are very small
compared to the values in generic quantum Hall setups 
which have magnetic fields and electron densities 
higher by about a factor $10^4$. Thus, for realization
one has to look for systems with high mobility at much smaller electron
densities or to make the geometric structures of the sample smaller, e.g.,
a factor 5 in linear dimensions yields a factor $25$ in the electron density 
and in the magnetic field.

An interesting alternative to standard semiconductors is the quantum
Hall effect in graphene. The relation between magnetic length $\ell_B$ and
magnetic field is the same \cite{brey06,abani07,delpl10,wang11c,stegm15}, but the relevant electron density $n$ is measured
relative to the semimetal so that small values are easily realized. 
Due to the density-of-states linear in energy one has $n\propto 
\epsilon_\text{F}^2$.
Furthermore, due to the perfect lattice structure a high mobility can
be expected. So the promising aim is to create non-trivial boundaries
with bays on the scale of $10$ to $1000$nm in graphene.

In conclusion, an experimentally realizable topological phase, the integer quantum Hall effect,
allows for tunable Fermi velocities if its edges are appropriately decorated. Gate voltages can 
serve as  control parameters for tuning. These findings should 
encourage further research
to realize such systems on the laboratory level to ultimately pave the way towards real devices.

As an outlook we want to emphasize that the presented finding can be
extended in several ways as has been done for lattice models \cite{malki17b}.
The detrimental effects of disorder can be included to study
the robustness of the observed effects. Such investigations will help 
to understand with which accuracy an experimental realization has to
be grown in order to be able to observe the predicted effects.
Without doubt, this constitutes an essential step toward applications.

Second, our findings can be extended 
to spinful models without conceptual difficulties. If the spin is subject to
spin-orbit coupling the chiral edge modes will generically become helical modes
which opens up the promising field for applications in spintronics, for instance
realizing switchable spin diodes. Thus, many tantalizing research projects lie ahead.

\section*{Acknowledgements}
We acknowledge useful discussions with Manfred Bayer, Axel Lorke, Bruce Normand, and
Dirk Reuter.

\paragraph{Funding information}
One of the authors (MM) gratefully acknowledge financial support by the Studienstiftung des deutschen Volkes. This work was also supported by the Deutsche Forschungsgemeinschaft and the
Russian Foundation of Basic Research in the International Collaborative
Research Center TRR 160.





\nolinenumbers

\end{document}